\documentstyle[prd,aps]{revtex}
\newcommand{\be}{\begin{equation}}
\newcommand{\ee}{\end{equation}}
\begin{document}
\draft
\title{The \v Cerenkov effect with massive photons}
\author{Miroslav Pardy}
\address{Department of Theoretical Physics and Astrophysics,
Masaryk University,
Kotl\'{a}\v{r}sk\'{a} 2, 611 37 Brno\\
Czech Republic\\
E-mail: pamir@physics.muni.cz}
\date{\today}
\maketitle
\begin{abstract}
Then equations of massive electrodynamics
are derived and the power spectrum formula for the \v{C}erenkov radiation
of massive photons is found. The \v Cerenkov power spectrum is determined
also for the two charge system. It is argued that the massive \v Cerenkov
effect can be observed in superconductive media, ionosphere plasma,
waveguides and in particle laboratories.
\end{abstract}
\hspace{3ex}

\baselineskip 11pt
\section{Introduction}

The possibility that photon may be massive particle has been treated
by many physicists. At the present time the great attention is devoted
to discussion of the mass of neutrino and its oscillations,
nevertheless theoretical problems with massive
photons is of the same importance. The established fact is that the massive
electrodynamics is a perfectly consistent classical and
quantum field theory [1].
In all respect the quantum version has the same status as the standard QED.
In this article we do not solve the radiative problems in sense of
[2], our goal is to determine the \v Cerenkov effect
of massive photons which is not in [2] analyzed.

In particle physics and quantum field theory [3-5] photon is defined
as a massless particle with spin 1. Its spin is along or opposite to its
motion. The massive photon as a neutral massive particle
is usually called vector boson.
There are other well known examples of massive
spin-1 particles. For instance neutral $\varrho$-meson, $\varphi$-meson and
$J/\psi$ particle, bosons $W^{\pm}$ and $Z^{0}$ in particle physics.

While massless photon is described by the Maxwell
Lagrangian, the massive photon is described by the Proca Lagrangian from
which the field equations follow. The massive electrodynamics can be
considered as a generalization of massless electrodynamics.
The well known area where the massive photon or boson plays
substantial role is
the theory  of superconductivity [3], plasma physics [6], waveguides and so on.
So, the physics of massive photon is meaningful and
and it means that also the \v Cerenkov effect with massive
photons is worthwile to investigate.

In order to be pedagogically clear, we treat
in section II the massive spin 0 quantum field theory and then
in section III the massive spin 1 field theory. In section IV
the massive Maxwell equations are derived.
In that section power spectral formula is derived
for the massive \v Cerenkov radiation for the situation
with one charge moving in a medium. In section V,
we derive the \v Cerenkov spectrum of massive photons
for the system of two charges moving in a medium.

\section{Massive spin 0 fields}

We begin with the massive spin 0 fields as the most simple illustration
how the source theory works [7].
The action or spin 0 particles is  according to source theory composed
from the scalar source $K(x)$ and propagator $\Delta_{+}$ in such
a way that it gives the correct probability condition for the
vacuum to vacuum amplitude. We show here that the action is

$$W(K) = \frac {1}{2}\*\int\,(dx)(dx')K(x)\Delta_{+}(x-x')K(x'),
\eqno(1)$$
and gives the right probability condition
$|\langle 0_{+}|0_{-}\rangle|^2 \leq 1$, where ($\hbar = 1$) [7,8]

$$\langle 0_{+}|0_{-}\rangle ^{K} = e^{iW(K)} \eqno(2)$$
is the basic formula of the Schwinger source theory with
$\langle 0_{+}|0_{-} \rangle$ being the vacuum to vacuum amplitude.

In order to prove that the quantity $\langle 0_{+}|0_{-} \rangle$ is really
the vacuum to vacuum amplitude it is necessary to know the
explicit form of the Green function $\Delta_{+}(x - x')$ which
satisfies to equation

$$\left(-\partial ^2 + m^2\right)\Delta_{+}(x - x') =
\delta(x - x'). \eqno(3)$$

From the last eq. follows that

$$\Delta_{+}(x-x') = \frac {1}{(-\partial ^2 + m^2)}
\int\; \frac {(dp)}{(2\pi)^4}e^{ip(x-x')}. \eqno(4)$$

The formula (4) is not unambiguos and it is necessary to specify it
by the $\varepsilon$-term, or,

$$\Delta_{+}(x-x') = \int\; \frac {(dp)}{(2\pi)^4}
\frac {e^{ip(x-x')}}{p^2 + m^2 -i\epsilon}; \quad
\epsilon \rightarrow 0_{+}. \eqno(5)$$

Now, let us proof that $|\langle 0_{+}|0_{-}\rangle|^2$ is the probability
of the persistence of vacuum. According to definition

$$\langle 0_{+}|0_{-}\rangle  = \exp\left\{
\frac {i}{2}\*\int\,(dx)(dx')K(x)\Delta_{+}(x-x')K(x')\right\}, \eqno(6)$$
or,

$$\langle 0_{+}|0_{-}\rangle  = \exp\left\{\frac {i}{2}\int\; (dx)(dx')
 \int\; \frac {(dp)}{(2\pi)^4} K(x)
\frac {e^{ip(x-x')}}{p^2 + m^2 -i\epsilon} K(x')\right\} \quad =$$

$$\exp\left\{\frac {i}{2} \int\; \frac {(dp)}{(2\pi)^4}
\frac {K(p)K(-p)}{p^2 + m^2 -i\epsilon}\right\} \quad =
\exp\left\{\frac {i}{2} \int\; \frac {(dp)}{(2\pi)^4}
\frac {|K(p)|^2}{p^2 + m^2 -i\epsilon}\right\} \eqno(7)$$
as a consequence of eq. (5) and $K^{*}(p) = K(-p)$.
Using the well known theorem

$$\frac {1}{x - i\varepsilon} = P\left(\frac {1}{x}\right) +
i\pi\delta(x);\quad \varepsilon \rightarrow 0,\eqno(8)$$
where $P$ denotes the principal value of integral we get the following
formula for the vacuum persistence:

$$|\langle 0_{+}|0_{-}\rangle|^2 = e^{-2{\rm Im}\;W} =
\exp\left\{- 2 \int\; \frac {(dp)}{(2\pi)^4}
\pi|K(p)|^2 \delta(p^2 + m^2)\right\}. \eqno(9)$$

Using

$$\delta(p^2 + m^2)\quad = \frac {1}{2({\bf p}^2 + m^2)^{1/2}}\*
\left\{\delta\left(p^{0} - ({\bf p}^2 + m^2)^{1/2}\right) +
\delta\left(p^{0} + ({\bf p}^2 + m^2)^{1/2}\right)\right\}, \eqno(10)$$
we get

$$2 \int\; \frac {(dp)}{(2\pi)^4}
\pi|K(p)|^2\delta(p^2 + m^2) = \int\; \frac {(d{\bf p})}{(2\pi)^3}\frac
{1}{2p^{0}}|K(p^{0},{\bf p})|^2 ,\eqno(11)$$
and then,

$$|\langle 0_{+}|0_{-}\rangle|^2 = \exp\left\{-\int\; d\omega_{p}|K(p)|^2
\right\}, \eqno(12)$$
where

$$d\omega_{p} = \frac {(d{\bf p})}{(2\pi)^3}\*\frac {1}{2p^0}\quad
p^0 = +({\bf p}^2 + m^2)^{1/2}. \eqno(13)$$

The expression  (12) shows that in the presence of the scalar source $K(x)$
the probability for vacuum to remain a vacuum is equal or less than 1.

Now, let us show the derivation of the field equation from the action
$W$ for the scalar field $\varphi$, where

$$W = \frac {1}{2} \int K\Delta_{+}K = \frac {1}{2}\int\varphi K =
\frac {1}{2}\int \varphi(-\partial^{2} + m^{2})\varphi =
\frac {1}{2}\int (\partial_{\mu}\varphi\partial^{\mu}\varphi +
m^{2}\varphi^{2}) = $$

$$2W - W = \int \varphi K - \frac {1}{2}[(\partial \varphi)^{2} +
m^{2}\varphi^{2}] =
\int\;(dx)\left[K(x)\varphi(x) +  {\cal L} (\varphi(x))
\right], \eqno(14)$$
with

$$ {\cal L}(\varphi(x)) = -\frac{1}{2}\left[
\partial_{\mu}\varphi\partial^{\mu}\varphi + m^2\varphi^2\right].
\eqno(15)$$

Let us put

$$\delta_{\varphi}W = 0,\eqno(16)$$
or,

$$\int\;\delta\varphi K~- \left[
\partial_{\mu}\varphi\partial^{\mu}\delta\varphi + m^2\varphi
\delta\varphi\right] = 0.\eqno(17)$$

After some modification we get

$$\int\;(dx)\left[K - (-\partial_{\mu}\partial^{\mu}\varphi + m^2\varphi)
\right]\delta\varphi = 0. \eqno(18)$$

As variable $\varphi$ is an arbitrary one the last integral is equal
to zero only if

$$\left(-\partial ^2 + m^2\right)\varphi(x) =  K(x),\eqno(19)$$
which is the Klein-Gordon equation with source $K(x)$ on the right side
of equation.
Now, let us derive the Proca equation for massive particles with spin 1
and generate the Maxwell equations for massive photons.

\section{Massive fields with spin 1}

We show the natural construction of the field of the particles with
spin 1. The derivation of the action for this massive spin 1 fields
is based on the modification of the derivation of spin 0 fields.

The relation

$$|\langle 0_{+}|0_{-}\rangle|^2 = \exp\left\{-2{\rm Im}\;W\right\}
 \leq 1 \eqno(20)$$
is postulated to be valid for all spin fields. Let us show here the
construction of action and field equations concerning spin one.

If spin zero particles and fields are described by the scalar source,
then a vector source denoted here as $J^{\mu}(x)$ can be considered
as a candidate for the description of the spin 1 fields and
particles. However, there exist some obstacles because source
$J^{\mu}(x)$ has four components and spin one particles have only
three spin possibilities. Nevertheless first, let us investigate by
analogy with the spin zero fields the following form of the action
for the unit spin fields:

$$W(J) = \frac {1}{2}\int\; (dx)(dx')
J^{\mu}(x)\Delta_{+}(x-x')J_{\mu}(x'). \eqno(21)$$

Then,

$$|\langle 0_{+}|0_{-}\rangle|^{2} = e^{iW}e^{iW^*} =
\exp\left\{-\int\; d\omega_{p}J^{*\mu}(p)J_{\mu}(p)\right\}. \eqno(22)$$

However,
$$J^{*\mu}(p)J_{\mu}(p) = |{\bf J}(p)|^2 - |J^{0}(p)|^2  \leq 0,\quad
{\rm or},\; > 0 \eqno(23)$$
and it means that the quantity defined by eq. (21) cannot be
considered as the probability of the persistence of vacuum.

The difficulty can be overcome by replacing the original form
$J^{*\mu}(x)J_{\mu}(x)$ by the following invariant structure:

$$J^{*\mu}(p)\left[g_{\mu\nu} + \frac
{1}{m^2}p_{\mu}p_{\nu}\right]J^{\nu}(p),
\eqno(24)$$
which can be with regard to its invariancy, determined in the rest frame
of the time-like vector $p^{\mu}$, where $p^{\mu} = (m, 0, 0, 0)$ in
the rest frame. Then, with $g_{\alpha\alpha} = (-1, 1, 1, 1)$ and
$g_{\mu\nu} = 0$  for $\mu \ne \nu$ we have

$$\bar g_{\mu\nu} = g_{\mu\nu} + \frac {1}{m^2}p_{\mu}p_{\nu} = \left\{
\begin{array}{ccc}
\delta_{kl}; & \mu = k; & \nu = l\\
0; & \mu = 0; & \nu = 0\\
0; & \mu = k; & \nu = 0
\end{array}
\right.
\eqno(25)$$
and

$$J^{*\mu}(p)\bar g_{\mu\nu}J^{\nu}(p) \equiv |{\bf J}|^2 \eqno(26)$$
and now the quantity $|\langle 0_{+}|0_{-}\rangle|^2$ can be
interpreted as the vacuum persistence probability.

At the same time $|{\bf J}|^2$ contains three independent source
components, transforming among themselves under spatial rotation, as
it is appropriate to unit spin.

After using eq. (24) it may be easy to get $W(J)$ in the space-time
representation
by the Fourier transformation, as it follows

$$W(J) = \frac {1}{2}\int (dx)(dx')
\left\{J_{\mu}(x)\Delta_{+}(x-x')J^{\mu}(x') + \frac {1}{m^2}
\partial_{\mu}J^{\mu}(x)\Delta_{+}(x-x')\partial_{\nu}'J^{\nu}(x')\right\}.
\eqno(27)$$

The field of spin one particles can be defined using the definition
of the test source $\delta J^{\mu}(x)$ by the relation

$$\delta W(J) = \int\;(dx) \delta J^{\mu}(x)\varphi_{\mu}(x), \eqno(28)$$
where $\varphi_{\mu}$ is the field of particles with spin 1. After
performing variation of the formula (27) and comparison with eq. (28)
we get the equation for field of spin 1 in the following form:

$$\varphi_{\mu}(x) = \int\; (dx')\Delta_{+}(x-x')J_{\mu}(x') - \frac {1}{m^2}
\partial_{\mu}\int\; (dx')\Delta_{+}(x-x')\partial_{\nu}'J^{\nu}(x').
\eqno(29)$$

The divergence of the vector field $\varphi_{\mu}(x)$ is given by the
relation

$$\partial_{\mu}\varphi^{\mu}(x) = \int\; (dx')\Delta_{+}(x-x')
\partial_{\mu}'J^{\mu}(x') -
\frac {1}{m^2}\partial^2 \int (dx')\Delta_{+}(x-x')
\partial_{\nu}'J^{\nu}(x') =
\frac {1}{m^2}\partial_{\mu}J^{\mu}(x), \eqno(30)$$
as a cosequence of eq. (5) and  relation

$$-\partial^2\Delta_{+} = \delta(x-x') - m^2\Delta_{+}. \eqno(31)$$.

Further, we have after applying operator $(-\partial ^2 + m^2)$ on
the equation (29) the following equations:

$$\left(-\partial ^2 + m^2\right)\varphi_{\mu}(x) =
J_{\mu}(x) - \frac {1}{m^2}\partial_{\mu}\partial_{\nu}J^{\nu}(x),
\eqno(32)$$

$$\left(-\partial ^2 + m^2\right)\varphi_{\mu}(x) +
\partial_{\mu}\partial_{\nu}\varphi^{\nu}(x) = J_{\mu}(x),
\eqno(33)$$
as a consequence of eq. (30).

It may be easy to cast the last equation into the following form

$$\partial^{\nu}G_{\mu\nu} + m^2\varphi_{\mu} = J_{\mu}, \eqno(34)$$
where

$$G_{\mu\nu}(x) = -G_{\nu\mu}(x) =
\partial_{\mu}\varphi_{\nu} - \partial_{\nu}\varphi_{\mu}. \eqno(35)$$

Identifying $G_{\mu\nu}$ with $F_{\mu\nu}$ of the electromagnetic
field we get instead of eq. (33) and eq. (34) so called
the Proca equation for the
electromagnetic field with the massive photon.

$$\left(-\partial ^2 + m^2\right)A_{\mu}(x) +
\partial_{\mu}\partial_{\nu}A^{\nu}(x) = J_{\mu}(x), \eqno(36)$$

$$\partial^{\nu}F_{\mu\nu} + m^2A_{\mu} = J_{\mu}, \eqno(37)$$

$$F_{\mu\nu}(x) = -F_{\nu\mu}(x) =
\partial_{\mu}A_{\nu} - \partial_{\nu}A_{\mu}. \eqno(38)$$

In case $m^{2} \ne 0$, we can put $\partial_{\mu}A^{\mu} = 0$ in order
to get:

$$\left(-\partial ^2 + m^2\right)A_{\mu}(x) = 0, \quad
\partial_{\mu}A^{\mu} = 0. \eqno(39)$$

The solution of the system (39) is the plane wave

$$ A_{\mu} = \varepsilon_{\mu}({\bf k})e^{ikx}, \quad k^{2} = -m^{2}
\eqno(40)$$
with $k\varepsilon({\bf k}) = 0$, which is precisely the correct definition
of a massive particle with spin 1. We will see
in the next section how to generalize this procedure
to the situation of the massive electrodynamics in dielectric
and magnetic media and then to apply
it to the determination of the massive \v Cerenkov radiation.

The equation (34) can be derived also from the action

$$W = \int (dx)\left(J^{\mu}(x)\varphi_{\mu}(x) +  {\cal L}
(\varphi(x))\right),
\eqno(41)$$
where

$$ {\cal L} = -\frac {1}{2}\left(\frac {1}{2}
(\partial^{\mu}\varphi^{\nu} - \partial^{\nu}\varphi^{\mu})
(\partial_{\mu}\varphi_{\nu} - \partial_{\nu}\varphi_{\mu}) +
m^{2}\varphi^{\mu}\varphi_{\mu}\right),\eqno(42)$$
where we have used the arrangement

$$\int (dx)\varphi^{\mu}(-\partial^{2})\varphi_{\mu} = \int (dx)
\partial^{\nu}\varphi^{\mu}\partial_{\nu}\varphi_{\mu} \eqno(43)$$
and

$$\int (dx)\varphi^{\mu}\partial_{\mu}\partial^{\nu}\varphi_{\nu} =
- \int (dx)\varphi^{\nu}\partial^{\mu}\partial_{\mu}\varphi_{\nu} =
- \int (dx)\varphi_{\mu}\partial^{\mu}\partial^{\nu}\varphi_{\nu}.
\eqno(44)$$

Using the last equation (44) we get the Lagrange function in the following
standard form:

$$ {\cal L} = -\frac{1}{2}
\left(\partial^{\nu}\varphi^{\mu}\partial_{\nu}\varphi_{\mu} -
(\partial_{\mu}\varphi^{\mu})^{2} + m^{2}\varphi^{\mu}\varphi_{\mu}\right).
\eqno(45)$$

If we use the A- and F-symbols, we receive from eq. (42) the Proca
Lagrangian

$$ {\cal L} = -\frac {1}{2}\left(\frac {1}{2}F^{\mu\nu}F_{\mu\nu} +
m^{2}A^{\mu}A_{\nu}\right),\eqno(46)$$
or,

$$ {\cal L} =
-\frac {1}{2}\left(\partial^{\nu}A^{\mu}\partial_{\nu}A_{\mu} -
(\partial_{\mu}A^{\mu})^{2} + m^{2}A^{\mu}A_{\mu}\right).
\eqno(47)$$

By variation of the corresponding Lagrangians for the massive field
with spin 1 we get evidently the massive Maxwell equations.

It is evident that the zero mass limit does not exist for
$\partial_{\mu}J^{\mu}(x) \not = 0$. In such a way we are forced to
redefine action $W(J)$. One of the possibilities is to put

$$\partial_{\mu}J^{\mu}(x) = mK(x) \eqno(48)$$
and identify $K(x)$ in the limit $m\rightarrow 0$ with the source of
massless spin zero particles. Since the zero mass particles with zero
spin are experimentally unknown in any event, we take $K(x) = 0$ and
we write

$$W_{[m=0]}(J) = \frac {1}{2}\int\; (dx)(dx')J_{\mu}(x)D_{+}(x-x')
J^{\mu}(x'), \eqno(49)$$
where
$$\partial_{\mu}J^{\mu}(x) = 0 \eqno(50)$$
and
$$D_{+}(x-x') = \Delta_{+}(x-x'; m=0). \eqno(51)$$

In case we want to work with electrodynamics in medium it is necessary
to involve such parameters as velocity of light $c$ magnetic permeability
$\mu$ and the dielectric constant $\varepsilon$. Then the corresponding
equations for electromagnetical potentials which are compatible with the
Maxwell equations are as follows [8]:

$$\left(\Delta-\frac{\mu\epsilon}{c^2}
\*\frac{\partial^2}{\partial\*t^2}\right)A^{\mu}
= \frac{\mu}{c}\left(g^{\mu\nu}+\frac{n^2-1}{n^2}\eta^\mu\*\eta^\nu\right)
J_{\nu},\eqno(52)$$
where the corresponding Lorentz gauge is defined in the Schwinger et al.
article in the following form

$$\partial_{\mu}A^{\mu} - (\mu\varepsilon-1)(\eta\partial)(\eta A) = 0,
\eqno(53)$$
where $\eta^{\mu} = (1,{\bf 0})$ is the unit timelike vector in the rest
frame of the medium.
The four-potentials are $A^{\mu}(\phi,{\bf A~})$ and
the four-current $J^{\mu}(c\varrho,{\bf J})$,
$n$ is the index of refraction of this medium.

The corresponding Green function $D_{+\mu\nu}$
in the $x$-representation is:

$$D_{+}^{\mu\nu}(x-x') = \frac{\mu}{c}\*
\left(g^{\mu\nu}+\frac{n^2-1}{n^2}\*
\eta^{\mu}\eta^{\nu}\right)\*D_{+}(x-x').\eqno(54)$$

$D_{+}(x-x')$ was derived by Schwinger et al. [8] as follows:

$$D_{+}(x-x') = \int \frac {(dk)}{(2\pi)^4}\*\frac {e^{ik(x-x')}}
{|{\bf k}^2| - n^2(k^0)^2 - i\epsilon},
\eqno(55)$$
Or,

$$D_{+}(x-x') = \frac {i}{c}\*\frac {1}{4\pi^2}\*\int_{0}^{\infty}d\omega
\frac {\sin\frac {n\omega}{c}\*|{\bf x}-{\bf x}'|}{|{\bf x}-{\bf x}'|}
\*e^{-i\omega|t-t'|}.
\eqno(56)$$

\section{Massive photon in electrodynamics and the \v Cerenkov effect}

The massive electrodynamics in medium can be constructed by generalization
of massless electrodynamics to the case with massive photon. In our case it
means that we replace only eq. (52) by the following one:

$$\left(\Delta-\frac{\mu\epsilon}{c^2}
\frac{\partial^2}{\partial\*t^2} + \frac {m^{2}c^{2}}{\hbar^{2}}
\right)A^{\mu}
= \frac{\mu}{c}\left(g^{\mu\nu}+\frac{n^2-1}{n^2}\eta^\mu\*\eta^\nu\right)
J_{\nu},\eqno(57)$$
where $m$ is mass of photon. The Lorentz gauge (53) is conserved also
in the massive situation.

In superconductiviy photon is a massive spin 1 particle
as a consequence of a broken symmetry of the
Landau-Ginzburg Lagrangian. The Meissner effect can be used as a
experimental demonstration that photon in a
superconductor is a massive particle.
In particle physics the situation is analogous to the situation in
superconductivity. The masses of particles are also generated by the broken
symmetry or in other words by the Higgs mechanism. Massive particles
with spin 1 form the analogue of the massive photon.

Kirzhnitz and Linde proposed a qualitative analysis wherein they
indicated that, as in the Ginzburg-Landau theory of superconductivity,
the Meissner effect can also be realized in the Weinberg model. Later,
it was shown that the Meissner
effect is  realizable in renormalizable gauge fields and also in the Weinberg
model [9].

We concentrate in this article to the  \v Cerenkov radiation
with massive photons.
The so called \v Cerenkov radiation was observed
experimentally first by \v{C}erenkov [10] and theoretically
explained by Tamm and Frank [11] in classical electrodynamics as
a shock wave resulting from a charged particle moving through a material
faster than the velocity of light in the material.
The source theory explanation was given by
Schwinger et al. [8] and the particle production by the \v{C}erenkov
mechanism was discussed by Pardy [12,13]. The \v Cerenkov effect at finite
temperature in source theory was discussed in [14,15] and
the \v Cerenkov effect
with radiative corrections, in electromagnetism and gravity
was analysed in [16,17].

We will investigate how the spectrum of the \v{C}erenkov radiation is
modified if we suppose the massive photons are generated instead
of massless photons.
The derived results form an analogue of the situation with massless photons.
According to [14--18], and with the analogy of the massless
photon propagator $D(k)$ in the momentum representation

$$D(k)  =  \frac {1}{|{\bf k}|^2-n^2(k^0)^2-i\epsilon}, \eqno(58) $$
the massive photon propagator is of the form
(here we introduce $\hbar$ and $c$):

$$D(k,m^{2}) =  \frac {1}
{|{\bf k}|^2-n^2(k^0)^2+\frac {m^2\*c^2}{\hbar^{2}}-i\epsilon},
\eqno(59)$$
where this propagator is derived from an assumption that the photon
energetical equation is

$$|{\bf k}|^2 - n^2(k^0)^2 = - \frac {m^2\*c^2}{\hbar^{2}},
\eqno(60)$$
where $n$ is the parameter of the medium and $m$ is mass of
photon in this medium.

From eq. (60) the dispersion law for the massive photons follows:
$$\omega = \frac {c}{n}\sqrt{k^{2} + \frac {m^{2}c^{2}}{\hbar^{2}}}.
\eqno(61)$$

Let us remark here that such dispersion law is valid not only for
the massive photon but also for electromagnetic field in waveguides
and electromagnetic field in ionosphere. It means that the corresponding
photons are also massive and the theory of massive photons is physically
meaningful. It means that also the \v Cerenkov radiation of
massive photons is physically meaningful and it is worthwile to study it.

The validity of eq. (60) can be verified using very simple idea
that for $n = 1$ the Einstein equation for mass and energy has to follow.
Putting ${\bf p} = \hbar{\bf k}, \quad \hbar k^{0} =
\hbar (\omega/c) = (E/c)$, we get the Einstein energetical equation

$$E^{2} = {\bf p}^{2}c^{2} + m^{2}c^{4}.
\eqno(62)$$

The propagator for the massive photon is then derived as

$$D_{+}(x-x',m^{2}) = \frac{i}{c}\frac{1}{4\pi^2}
\int_{0}^{\infty}\,d\omega\,\frac{\sin[\frac{n^2\omega^2}{c^2}-
\frac {m^2\*c^2}{\hbar^{2}}]^{1/2}
|{\bf x}-{\bf x}'|}{|{\bf x}-{\bf x}'|}\*e^{-i\omega\*|t-t'|}.
\eqno(63)$$

The function (63) differs from the the original function
$D_{+}$ by the factor

$$\left(\frac {\omega^2\*n^2}{c^2} -
\frac {m^2\*c^2}{\hbar^{2}}\right)^{1/2}.
\eqno(64)$$

From eq. (56) and (63) the potentials generated by the massless or massive
photons respectively follow. In case of the massless photon, the potential is
according to Schwinger defined by the formula:

$$V({\bf x} - {\bf x'}) = \int_{-\infty}^{\infty}d\tau
D_{+}({\bf x - \bf x}',\tau) =
\int_{-\infty}^{\infty}d\tau \left\{
\frac {i}{c}\*\frac {1}{4\pi^2}\*\int_{0}^{\infty}d\omega \frac {\sin
\frac {n\omega}{c}\*|{\bf x}-{\bf x}'|}{|{\bf x}-{\bf x}'|}\*e^{-i\omega|\tau|}
\right\}.
\eqno(65)$$

The $\tau$-integral can be evaluated using the mathematical formula

$$\int_{-\infty}^{\infty}\,d\tau\, e^{-i\omega|\tau|} = \frac {2}{i\omega}
\eqno(66)$$
and the $\omega$-integral can be evaluated using the formula

$$\int_{0}^{\infty}\frac {\sin ax}{x}dx = \frac {\pi}{2}, \quad {\rm for}
\quad a>0.
\eqno(67)$$

After using eqs. (66) and (67), we get

$$V({\bf x} - {\bf x'}) =\frac {1}{c} \frac {1}{4\pi}
\frac {1}{|{\bf x} - {\bf x}'|}.
\eqno(68)$$

In case of the massive photon, the mathematical determination of potential
is the analogical to the massless situation only with the difference we use
the propagator (63) and the table integral [19]

$$\int_{0}^{\infty}\frac {dx}{x}\sin\left(p\sqrt{x^{2}-u^{2}}\right) =
\frac {\pi}{2}e^{-pu}.
\eqno(69)$$

Using this integral we get that the potential generated by the massive
photons is

$$V({\bf x} - {\bf x'},m^{2}) = \frac {1}{c}\frac {1}{4\pi}
\frac {\exp{\left\{-\frac {mcn}{\hbar}|{\bf x} - {\bf x'}|\right\}}}
{|{\bf x} - {\bf x}'|}.
\eqno(70)$$

If we compare the potentials concerning massive and massless photons, we can
deduce that also \v Cerenkov radiation with masive photons can be generated.
So, the determination of the
\v Cerenkov effect with massive photons is physically meaningful.

In case of the massive electromagnetic field in the medium,
the action $W$ is given by the following formula:

$$W = \frac{1}{2c^2}\*\int\,(dx)(dx')J^{\mu}(x)D_{+\mu\nu}(x-x',m^{2})
J^{\nu}(x'),
\eqno(71)$$
where

$$D_{+}^{\mu\nu} = \frac{\mu}{c}[g^{\mu\nu} +
(1-n^{-2})\eta^{\mu}\eta^{\nu}]\*D_{+}(x-x',m^{2}),
\eqno(72)$$
where $\eta^{\mu}\, \equiv \, (1,{\bf 0})$, $J^{\mu}\, \equiv \,(c\varrho,
{\bf J})$ is the conserved current, $\mu$ is the magnetic permeability of
the medium, $\epsilon$ is the dielectric constant od the medium and
$n=\sqrt{\epsilon\mu}$ is the index of refraction of the medium.

The probability of the persistence of vacuum follows from the vacuum
amplitude (2) in the following form:

$$|\langle 0_{+}|0_{-} \rangle|^2 = e^{-\frac{2}{\hbar}\*\rm Im\*W},
\eqno(73)$$
where ${\rm Im}\; W$ is the basis for the definition of the spectral
function $P(\omega,t)$ as follows:

$$-\frac{2}{\hbar}\*{\rm Im}\*W \;\stackrel{d}{=} \;
-\,\int\,dtd\omega\frac{P(\omega,t)}{\hbar \omega}.
\eqno(74)$$

Now, if we insert eq. (72) into eq. (71), we get
after extracting $P(\omega,t)$ the following general expression
for this spectral function:

$$P(\omega,t) = -\frac{\omega}{4\pi^2}\*\frac{\mu}{n^2}\*\int\,d{\bf x}
d{\bf x}'dt'\;
\left[\frac{\sin[\frac{n^2\omega^2}{c^2}-
\frac {m^2\*c^2}{\hbar^{2}}]^{1/2}]|{\bf x}-{\bf x}'|}
{|{\bf x}-{\bf x}'|}\right]\;\times $$

$$\cos[\omega\*(t-t')]\*[\varrho({\bf x},t)\varrho({\bf x}',t')
- \frac{n^2}{c^2}\*{\bf J}({\bf x},t)\cdot{\bf J}({\bf x}',t')].
\eqno(75)$$

Now, let us apply the formula (75) in order to get the
\v{C}erenkov distribution of massive photons. The \v{C}erenkov
radiation is produced by charged particle of charge $Q$ moving
at a constant velocity ${\bf v}$.  In such a way we can
write for the charge density and for the current density:

$$\varrho = Q\*\delta\*({\bf x}-{\bf v}t), \hspace{7mm}
{\bf J} = Q{\bf v}\delta\*({\bf x}-{\bf v}t).
\eqno(76)$$

After insertion of eq. (76) into eq. (75), we get ($v = |{\bf v}|$).

$$P(\omega,t) = \frac{Q^2}{4\pi^2}\*\frac{v\mu \omega}{c^2}\*
\left(1-\frac{1}{n^2\beta^2}\right)
\int_{\infty}^{\infty}\*\frac{d\tau}{\tau}\*
\sin\left(\left[\frac{n^2\omega^2}{c^2}-\frac {m^2\*c^2}{\hbar^{2}}
\right]^{1/2}\*
v\tau\right)\* \cos\*\omega\tau,
\eqno(77)$$
where we have put $\tau = t'-t, \beta = v/c$.

For $P(\omega,t)$, the situation leads to evaluation of the $\tau$-integral.
For this integral we have:

$$\int_{-\infty}^{\infty}\,\frac{d\tau}{\tau}
\*\sin\left(\left[\frac{n^2\omega^2}
{c^2} -\frac {c^2}{m^2}\right]^{1/2}
\*v\tau\right)\*\cos\omega\tau =
\left\{
\begin{array}{ll}
\pi, & 0<m^2<\frac{\omega^{2}}{c^2\*v^2}\*(n^2\beta^2 - 1)\\
0, &  m^2>\frac{\omega^{2}}{c^2\*v^2}(n^2\beta^2 -1).
\end{array}
\right.
\eqno(78)$$

From eq. (78) immediately follows that $m^2>0$ implies the \v{C}erenkov
threshold $n\beta >1$. From eq. (77) and (78) we get the spectral
formula of the \v{C}erenkov radiation of massive photons
in the form:

$$P(\omega,t) = \frac{Q^2}{4\pi}\*\frac{v \omega\mu}{c^2}\*
\left(1-\frac{1}{n^2\beta^2}\right)
\eqno(79)$$
for

$$\omega > \frac {mcv}{\hbar}\frac {1}{\sqrt{n^2\beta^2 - 1}}>0,
\eqno(80)$$
and $P(\omega,t) = 0$ for

$$\omega<\frac {mcv}{\hbar}\frac {1}{\sqrt{n^2\beta^2 -1}}.
\eqno(81)$$

Using the dispersion law (61) we can write the power spectrum $P(\omega)$ as
a function dependent on $k^{2}$. Then,

$$P(k^{2}) = \frac{Q^2}{4\pi}\*\frac{v \mu}{nc}\* \sqrt{k^{2} +
\frac{m^{2}c^{2}}{\hbar^{2}}}
\left(1-\frac{1}{n^2\beta^2}\right) ; \quad  n\beta > 1
\eqno(82)$$
and $P(\omega,t) = 0$ for $n\beta < 1$

The most simple way how to get the angle $\Theta$ between vectors
${\bf k}$ and ${\bf p}$
is the use the conservation laws for an energy and momentum.

$$E - \hbar\omega = E',\eqno(83)$$

$${\bf p} - \hbar{\bf k} = {\bf p'},\eqno(84)$$
where $E$ and $E'$ are energies of a moving particle before and after act of
emission of a photon  with energy $\hbar\omega$ and momentum $\hbar{\bf k}$,
and ${\bf p}$ and ${\bf p'}$ are momenta of the particle
before and after emission of the same photon.

If we raise the equations  (83) and (84) to the second power and
take the difference of these quadratic equations,  we can extract
the $\cos\Theta$ in the
form:

$$\cos\Theta = \frac {1}{n\beta}\left(1 + \frac {m^{2}c^{2}}{\hbar^{2}k^{2}}
\right)^{1/2} + \frac {\hbar k}{2p}\left(1 - \frac {1}{n^{2}}\right) -
\frac {m^{2}c^{2}}{2n^{2}p\hbar k},
\eqno(85)$$
which has the correct massless limit. The massless limit also gives
the sense of the parameter $n$ which is introduced in the massive
situation.We also observe that while in the massless situation the angle
of emission depends only on $n\beta$, in case of massive situation
it depends also on the wave vector $k$. It means that the emission of the
massive photons are emitted by the \v Cerenkov mechanism
in all space directions.
So, in experiment the \v Cerenkov production of massive photons can be
strictly distinquished from the \v Cerenkov production of massless photons
or from the hard production of spin 1 massive particles.

\section{The \v Cerenkov radiation of the two-charge system}

Instead of considering the \v Cerenkov radiation of motion of one
charge, we here consider the system of two equal charges
$Q$ with the constant mutual distance $a = |{\bf a}|$
moving with velocity ${\bf v}$ in dielectric medium. This text is an
analogue of the text in [20]. In this
situation the charge and the current densities for this system
are given by the by the following equations:

$$\varrho = Q[\delta({\bf x} - {\bf v}t) + \delta({\bf x} -{\bf a} -
{\bf v}t)]\eqno(86)$$

$${\bf J} = Q\*{\bf v}[\delta({\bf x}-{\bf v}t) + \delta({\bf x} - {\bf a}
- {\bf v}t)].\eqno(87)$$
where ${\bf a}$ is the vector going from the left charge to right charge
with the length of $ a = |{\bf a}|$ in the system $S$.

Let us suppose that ${\bf v} \parallel {\bf a} \parallel x$.
Then, after insertion of eq. (86) and (87)
into eq. (75), putting $\tau = t' - t$, and $\beta = v/c$,
where $v = |{\bf v}|$, we get
instead of the formula (75) the following relation:

$$P(\omega,t) = 2P_{1}(\omega,t) + P_{2}(\omega,t) + P_{3}(\omega,t),
\eqno(88)$$
where

$$P_{1}(\omega,t) = \frac{1}{4\pi^2}\frac{Q^{2}\mu\omega}{c^{2}}\*v
\left[1 -\frac{1}{n^2\beta^{2}}\right]
\int_{-\infty}^{\infty}\,d\tau
\frac{\sin\left(\left[\frac{n^2\omega^2}
{c^2} -\frac {m^2c^{2}}{\hbar^2}\right]^{1/2}
\*v\tau\right)}{\tau}\cos \omega\tau \eqno(89)$$

$$P_{2}(\omega,t) = \frac{1}{4\pi^2}\frac{Q^{2}\mu\omega}{c^{2}}v
\left[1 -\frac{1}{n^2\beta^{2}}\right]
\int_{-\infty}^{\infty}\,d\tau
\frac{\sin\left(\left[\frac{n^2\omega^2}
{c^2} -\frac {m^2c^{2}}{\hbar^2}\right]^{1/2}
|\frac {a}{v} + \tau|\right)}
{|\frac {a}{v} + \tau|}\cos\omega\tau \eqno(90)$$

$$P_{3}(\omega,t) = \frac{1}{4\pi^2}\frac{Q^{2}\mu\omega}{c^{2}}\*v
\left[1 -\frac{1}{n^2\beta^{2}}\right]
\int_{-\infty}^{\infty}\,d\tau
\frac{\sin\left(\left[\frac{n^2\omega^2}
{c^2} -\frac {m^{2}c^2}{\hbar^2}\right]^{1/2}
|\frac {a}{v} - \tau|\right)}
{|\frac {a}{v} - \tau|}\cos\omega\tau .\eqno(91)$$

The formula (89) contains the known integral:

$$J _{1} = \int_{-\infty}^{\infty}\,d\tau
\frac{\left(\left[\frac{n^2\omega^2}
{c^2} -\frac {m^{2}c^2}{\hbar^2}\right]^{1/2}
\*v\tau\right)}{\tau}\*
\cos \omega\tau = \left\{
\begin{array}{ll}
\pi; & n\beta > 1\\
0; & n\beta < 1
\end{array}
\right. .
\eqno(92)$$

Formulae (90) and (91) contain  the following integrals:

$$J_{2} = \int_{-\infty}^{\infty}\,d\tau\frac{
\left(\left[\frac{n^2\omega^2}
{c^2} -\frac {m^{2}c^2}{\hbar^2}\right]^{1/2}
|\frac {a}{v} + \tau|\right)}
{|\frac {a}{v} + \tau|}\cos\omega\tau \eqno(93)$$
and

$$J_{3} = \int_{-\infty}^{\infty}\,d\tau
\frac{\left(\left[\frac{n^2\omega^2}
{c^2} -\frac {m^{2}c^2}{\hbar^2}\right]^{1/2}
|\frac {a}{v} - \tau|\right)}
{|\frac {a}{v} - \tau|}\cos\omega\tau .\eqno(94)$$

Using the integral (92) we finally get the power spectral
formula $P_{1}$ of the produced photons:

$$P_{1}(\omega,t) = \frac{Q^2}{4\pi}\*\frac{\mu\omega}{c^{2}}
v\left[1 - \frac{1}{n^2\beta^{2}}\right];\quad n\beta>1 \eqno(95)$$
and

$$P_{1}(\omega,t) = 0;\quad n\beta < 1.\eqno(96)$$

Successive transformations

$$\frac {a}{v} + \tau = T,\quad \frac {a}{v} - \tau = T ,\eqno(97)$$
generates, after evaluations of the corresponding integrals $J_{2}, J_{3}$,
the corresponding spectral formulas $P_{2}, P_{3}$:

$$P_{2}(\omega,t) = \frac{Q^2}{4\pi}\*\frac{\mu\omega}{c^{2}}
\cos\left(\frac {\omega a}{v}\right)
v\left[1 - \frac{1}{n^2\beta^{2}}\right] = P_{3}; \quad n\beta>1
\eqno(98)$$
and

$$P_{2}(\omega,t) = P_{3}(\omega,t) = 0;\quad n\beta < 1 .
\eqno(99)$$

The sum of the partial spectral formula form the total radiation
emitted by the \v Cerenkov mechanism of the two-charge system. Using eq.
(61) we get final results in the following form:

$$P(k^{2},t) = 2(P_{1} + P_{2}) =
\frac {Q^{2}}{\pi}
\frac{\mu v}{c n}\sqrt{k^{2} + \frac {m^{2}c^{2}}{\hbar^{2}}}
\cos^{2}\left(\frac {a c}{2vn}
\sqrt{k^{2} + \frac {m^{2}c^{2}}{\hbar^{2}}}\right)
\left[1 - \frac{1}{n^2\beta^{2}}\right]; \quad n\beta>1 \eqno(100)$$
and

$$P(k^{2},t) = 0;\quad n\beta < 1 .\eqno(101)$$

\section{Discussion}

The distribution of massive photons generated by the \v{C}erenkov radiation
is derived here to our knowledge in the framework of the source theory
for the first time and there is no conventional derivation
of this effect in QED. As this effect
was not discussed in physical literature, we fill up the gap by this
article.

The velocity of the charged projectile which generates the massless
\v Cerenkov radiation can be considered during the process of radiation
constant because the energy loss due to radiative process is small.
However, in case of massive \v Cerenkov effect the energy loss of the
projectile may be large which means the projectile is strongly deccelerated.
It means the duration of the generation of massive photons is very short.
The velocity can be considered
constant only in case of very energetical and heavy charged projectile.

From the theoretical point of view, we used the massive electrodynamics which
is only the generalization of the massless electrodynamics.
So, our derivation of the \v Cerenkov radiation of massive photons
can be considered also as a generalization of the situation with the massless
photons.

The theory of the \v Cerenkov radiation of massive photons concerns
the photons not only in superconductive medium but also plasma medium in
electron gas, ionosphere medium or photons in waveguides.
The possibility of the
existence of the massive photons in neutron stars
is discussed by Voskresensky et al. [21]. The bosons
$W^{\pm}$ and $Z^{0}$ are also massive and it means that the generalization
of our approach to the situation in the standard model is evidently
feasible. Similarly, the generation of vectors mesons
$\rho, \varphi,J/\psi$ by the \v Cerenkov mechanism may be possible.
Probably, they can be
generated in a such nuclear medium where they play role of mediators of
nuclear forces.

The \v Cerenkov effect with massive photons can be in the experiment
strongly distinquished from the classical effect because the
emission of massive photons is distributed in all space directions.

We hope that with regard to the situation
in physics of superconductivity,
plasma physics, physics of ionosphere, waveguide physics,
particle physics,
where massive photons are present, sooner or
later \v Cerenkov effect with massive photons will be observed and the
theory presented in our article confirmed.

\end{document}